\documentclass[conference, a4paper]{IEEEtran}
\IEEEoverridecommandlockouts
\usepackage{cite}
\usepackage[left=1.62cm,right=1.62cm,top=1.9cm]{geometry}

\usepackage{amsmath,amssymb,amsfonts}
\usepackage{mdwmath}
\usepackage{algorithm, algorithmic}
\usepackage{graphicx}
\usepackage{mathabx}
\usepackage{mdwtab}
\usepackage{textcomp}
\usepackage{xcolor}
\usepackage{booktabs}
\usepackage{graphicx}
\usepackage{enumitem}

\usepackage[explicit]{titlesec}
\usepackage{subfigure}
\usepackage[numbers,sort&compress]{natbib} 
\def\BibTeX{{\rm B\kern-.05em{\sc i\kern-.025em b}\kern-.08em
    T\kern-.1667em\lower.7ex\hbox{E}\kern-.125emX}}
\IEEEoverridecommandlockouts
\IEEEpubid{\makebox[\columnwidth]{979-8-3503-4693-0/23/\$31.00~\copyright~2023 IEEE \hfill} \hspace{\columnsep}\makebox[\columnwidth]{ }}
\begin{document}
\titlespacing{\subsection}{0pt}{0.1ex plus 1ex minus .1ex}{0.2ex plus .2ex}
\titlespacing{\section}{0pt}{0.1ex plus 1ex minus .1ex}{0.2ex plus .2ex}

\title{Federated Learning-based Vehicle Trajectory Prediction against Cyberattacks\\
}

\author{\IEEEauthorblockN{Zhe Wang\IEEEauthorrefmark{1}, Tingkai Yan\IEEEauthorrefmark{3}}
\IEEEauthorblockA{King's College London\IEEEauthorrefmark{1} Imperial College London.\IEEEauthorrefmark{3} 
\\
tylor.wang@kcl.ac.uk, tky22@ic.ac.uk}}

\maketitle

\begin{abstract}
With the development of the Internet of Vehicles (IoV), vehicle wireless communication poses serious cybersecurity challenges. Faulty information, such as fake vehicle positions and speeds sent by surrounding vehicles, could cause vehicle collisions, traffic jams, and even casualties. Additionally, private vehicle data leakages, such as vehicle trajectory and user account information, may damage user property and security. Therefore, achieving a cyberattack-defense scheme in the IoV system with faulty data saturation is necessary. This paper proposes a Federated Learning-based Vehicle Trajectory Prediction Algorithm against Cyberattacks (FL-TP) to address the above problems. The FL-TP is intensively trained and tested using a publicly available Vehicular Reference Misbehavior (VeReMi) dataset with five types of cyberattacks: constant, constant offset, random, random offset, and eventual stop. The results show that the proposed FL-TP algorithm can improve cyberattack detection and trajectory prediction by up to 6.99\% and 54.86\%, respectively, under the maximum cyberattack permeability scenarios compared with benchmark methods.
\end{abstract}

\begin{IEEEkeywords}
Internet of vehicles, federated learning, cyberattack, trajectory prediction.
\end{IEEEkeywords}

\section{Introduction}
The development of communication networks has caught the attention of researchers regarding the impact of cyberattacks on future Advanced Driver Assistance Systems (ADAS) and autonomous driving technology systems \cite{naseer2019swarm,9650522}. Autonomous vehicles (AVs) must choose a safe path through traffic after perceiving and predicting the trajectories of surrounding vehicles. While this may be a common task for experienced human drivers, AVs may struggle when faced with wireless communication cyberattacks, such as replay attacks and random offset attacks \cite{9839627}. These attacks can flood the network with bogus messages, leading to collisions with legitimate vehicles. Moreover, the raw data of vehicles often includes drivers' privacy information, such as user identity, account passwords, and start, real-time, and destination locations of AVs.
It may cause property losses or causality of users once being disclosed to attackers through the process of Vehicle to Vehicle (V2V) data trading
Disclosure of driver's privacy data through the process of Vehicle to Vehicle (V2V) data trading \cite{guler2021framework} and knowledge exchanging \cite{alvi2022utility} can lead to property losses or harm to users, which is another problem that needs to be addressed.

Existing research has shifted focus from trajectory prediction algorithms based on fixed mathematical formulas to neural network technology such as computer vision, dynamic Bayesian networks, and Long Short-Term Memory (LSTM) models \cite{jiang2022multi}. With appropriate network layers, learning rate, activation function, and other hyperparameters, these neural network methods have significantly reduced trajectory prediction errors. However, state-of-the-art vehicle trajectory prediction algorithms have yet to fully explore previous challenges, such as cyberattacks and user privacy. In particular, no adaptation or discussion of trajectory prediction algorithms has been implemented for cyberattacks of different degrees. This work attempts to fill this gap by proposing FL-TP to address the above problems. The contributions of this paper can be summarized as follows:
\begin{figure*}[t] 
\setlength{\belowcaptionskip}{-3cm}   
\centering
\includegraphics[width=0.9\textwidth]{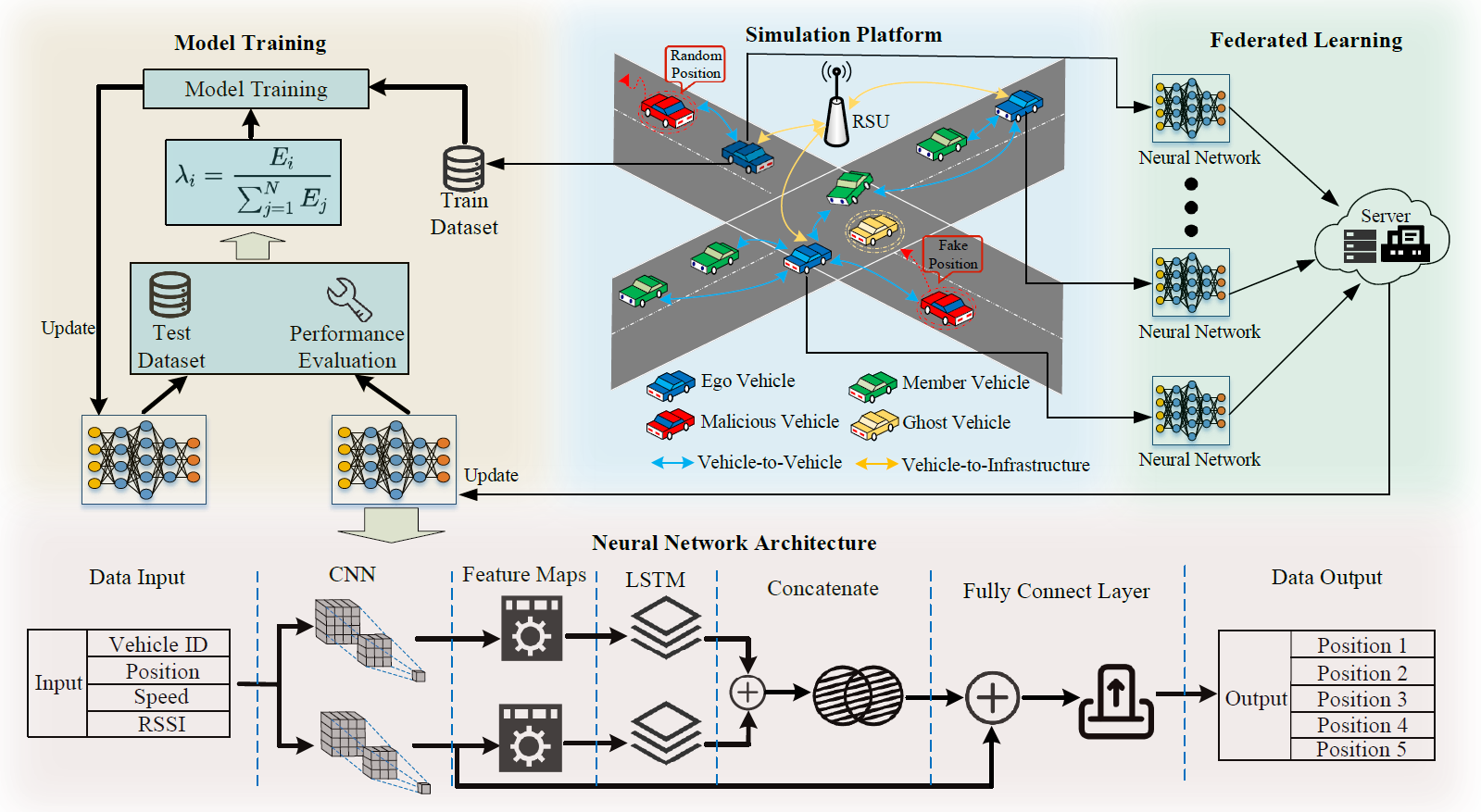}
\caption{ Scheme of the proposed FL-TP.}
\label{AlgoritmScheme}
\end{figure*}

\begin{enumerate}[itemsep=0pt]
\item  We propose an FL-TP algorithm based on the FL framework and LSTM network to enable vehicles to identify cyberattacks and efficiently predict the trajectories of surrounding vehicles under different cyberattack levels.

\item We propose a Model-robust Estimation (MrE) algorithm to calculate model aggregation weights while improving the effectiveness of FL in the global model aggregation process.

\item We perform various experiments in the IoV system to verify the FL-TP algorithm's trajectory prediction and attack detection performance on the VeReMi dataset.

\end{enumerate} 

The rest of this paper is organized as follows: Section II presents the FL-TP algorithm scheme. In section III, we provide details on the dataset, simulation setup, and experiment performance. In section IV, we detail the proposed MrE algorithm. In section V, we present the experiment performance of the proposed algorithm. Finally, section VI concludes this paper.

\section{related Works}
This section provides a review of related works in the IoV, including the progress of vehicle trajectory prediction algorithms, cyberattack detection, and federated learning.


\textbf{Vehicle trajectory prediction:} Recent advances in vehicle trajectory prediction have explored methods using deep neural networks \cite{yuan2023hierarchical}. For instance, Karimzadeh et al. propose an automated framework to predict future trajectories and traffic flow in urban areas without human interventions by integrating Reinforcement Learning (RL) and Transfer Learning (TL) \cite{9417511}.
Xing et al. propose a personalized Joint Time Series Modeling (JTSM) method based on the LSTM and Recurrent Neural Network model (RNN) to predict the front vehicle trajectories, building on the advances in LSTM Networks \cite{8933492}.

\textbf{Cyberattack detection:}
Researchers have started utilizing Deep Learning (DL) methods to address the attack detection problem with the availability of IoV cyberattack datasets, including the VeRemi, DARPA, and ISCX datasets
\cite{belenko2018synthetic}. For example, Khoa et al. proposed a novel collaborative machine learning-based intrusion detection system that can efficiently detect and prevent cyberattacks in the internet of things \cite{9120761}.

\textbf{Federated learning:}
As a distributed machine learning method, FL provides an efficient framework for addressing data island issues, preventing local data privacy leakage, and facilitating data collection from different terminals to form a large-scale knowledge exchange. For instance, Wang et al. proposed the Swarm Federated Deep Learning framework (IoV-SFDL) for the IoV system, which integrates blockchain technology and swarm learning into the FL framework. This integration aims to achieve higher security and performance in distributed systems for IoV scenarios \cite{wang2021credibility}.

\section{Proposed method}
The overview of the proposed FL-TP algorithm is shown in Fig. \ref{AlgoritmScheme}. The system model and problem formation will be introduced in this section.

\subsection{System model}
This paper considers a multi-vehicle broadcast scenario where the set of vehicles is denoted by $n={1, \cdots, N}$. Each vehicle is equipped with onboard radar, GPS, and Vehicular Ad-Hoc Network (VANet) communication equipment to obtain the position, speed, and real-time communication data from surrounding vehicles (SVs) $\{SV_{1}, SV_{2}, \cdots, SV_{M}\}$. To ensure data transmission efficiency, each SV must periodically group cast information to the ego vehicle. While receiving the message, each AV also obtains the Received Signal Strength Indicator (RSSI) of the message, which is defined as:
\begin{equation}
RSSI({dbm})=10*\log _{10}\frac{P(m w)}{1(m w)},
\end{equation}
where $P(mw)$ refers to the power of the received signal, measured in milliwatts, and is generally used to evaluate the link quality in wireless communication.
SVs can be divided into Member Vehicles (MeVs) and Malicious Vehicles (MaVs). MeVs are vehicles that periodically send accurate information to the ego vehicle, while MaVs are attack vehicles that periodically send bogus information. The paper defines five types of cyberattacks:

\textbf{Constant attack:}
This type of attack is also known as the ghost vehicle attack. In this attack, MaVs continuously broadcast false location information to the ego vehicle. By receiving these bogus messages, the ego vehicle detects a virtual vehicle on the initially empty road. The location data transmitted is defined as:
\begin{equation}
Loc_{constant}(t)=\{X_{f}, Y_{f}\},
\end{equation}
where $(X_{f}, Y_{f})$ refers to a fixed point within the region of interest that does not change with time.
\vspace{0.05cm}

\textbf{Constant and random offset attacks:} These two types of MaVs add fixed or random offsets to the real speed and location data before transmitting messages. As a result of these attacks, the ego vehicle may incorrectly estimate the speed of SVs. The location data transmitted during these attacks is defined as:
\begin{small} 
\begin{equation}
Loc_{offset}(t)=\left\{\begin{array}{ll}
(X_{t}+\Delta x_{f},Y_{t}+\Delta y_{f}) & Constant \\
(X_{t}+\Delta x_{r},Y_{t}+\Delta y_{r}) & Random ,
\end{array}\right .
\end{equation}
\end{small} 
where, $\Delta x_{f}$ and $\Delta y_{f}$ represent the fixed offset added by the MaVs based on their actual position data, and $\Delta x_{r}$ and $\Delta y_{r}$ represent the random offset. The MaVs select offset values based on the specific type of cyberattack.


\begin{figure}[t]
\setlength{\belowcaptionskip}{-3cm}   
\centering
\includegraphics[width=8.5cm]{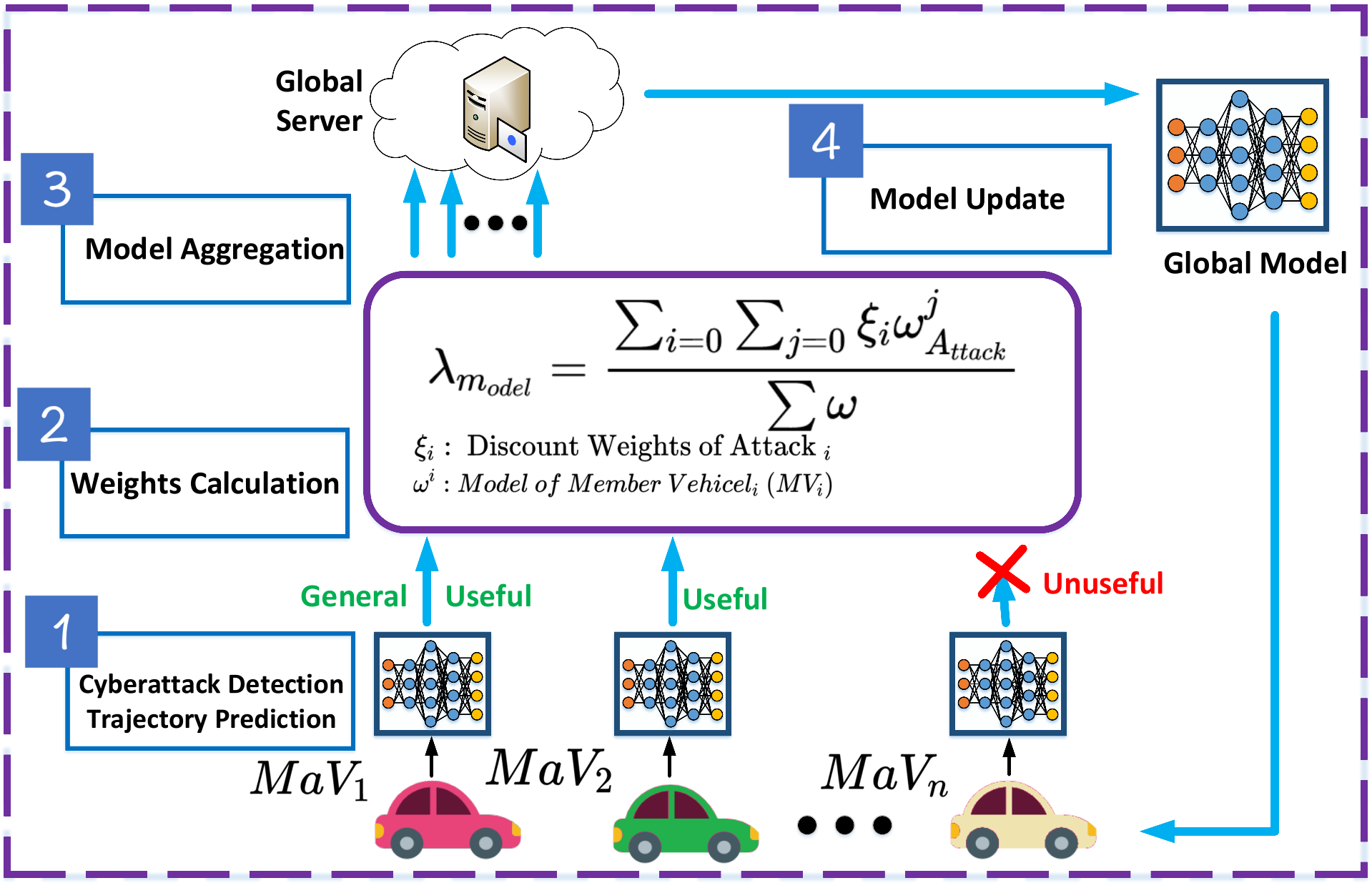}
\caption{Federated learning in FL-TP process}
\label{FDL}
\end{figure}

\textbf{Eventual stop attack:}
This kind of attack is also called the brake attack, and it is the most common and harmful attack in traffic scenarios. MaVs will broadcast sudden "brake to stop" messages while driving, causing ego vehicles to brake after receiving such faulty messages. This may cause traffic congestion and even accidents. The location data transmitted for the eventual stop attack is defined as:
\begin{equation}
Loc_{stop}(t)=\left\{\begin{array}{ll}
\left(X_{t}, Y_{t}\right) & P_{1} \\
\left(X_{t-1}, Y_{t-1}\right) & P_{2}
\end{array} \quad \sum P_{i}=1\right..
\end{equation}
The location data transmitted for the emergency stop attack is defined as $(X_t, Y_t)$, where $(X_{t-1}, Y_{t-1})$ represents the vehicle's location in the previous time step. The MaVs have a probability of $P_1$ to transmit their actual position coordinates, while there is a probability of $P_2$ to transmit the position coordinates of the previous time step to broadcast false emergency stop information to SVs.

\textbf{Random attack:}
In this kind of cyberattack, MaVs transmit a random position uniformly inside the area of interest. Although the random attack is easily recognizable, the message conveyed is meaningless to the entire traffic system and may cause serious traffic accidents. The transmitted location data is defined as:
\begin{equation}
Loc_{random}(t)=\left\{X_{r}, Y_{r}\right\} \quad X_{r}, Y_{r} \in[R, R],
\end{equation}
where $(X_{r}, Y_{r})$ refers to a random point in the area of interest, and $R$ refers to the width and height of the interest area of experiment.

The transmission delay of the data package in V2V communication is also considered in this paper for simulation. The time step when the vehicle receives the data packet $T_{rev}$ is defined as follows:
\begin{equation}
T_{rev}=T_{snd}+\frac{Distance}{Spd_{msg}},
\end{equation}
where the parameter $Distance$ refers to the euclidean distance between the sending vehicle and receiving vehicle, and $Spd_{msg}$ refers to the speed of message propagation. For easier understanding, $Spd_{msg}$ is defined as equal to the speed of light, which is also the same as the initial setting in the SUMO simulator.

\subsection{Problem formation}
This paper aims to predict the future trajectory of the target vehicle under various cyberattack levels. However, since the region of interest span is approximately $10$ km by $10$ km, the longitudinal and latitude positions can become quite large. Therefore, it is necessary to normalize the data before training the LSTM model to ensure that the model can converge quickly. To minimize the trajectory prediction error of each vehicle model for surrounding vehicles in the simulation environment, the input parameter $X$ includes the latitude and longitude coordinates and speed of SVs, the distance and speed difference between SVs and the ego vehicle, and the RSSI of the received data packet. $X$ is defined as:
\begin{equation}
\begin{aligned}
X=\{&Loc_x, Loc_y,Spd_x, Spd_y,disChg_x,\\&disChg_y,SpdChg_x,SpdChg_y,RSSI\}.
\end{aligned}
\end{equation}

The onboard base LSTM model has two tasks while driving: SVs trajectory prediction and cyberattack detection. Therefore, the label data $Y$ includes the trajectory data of SVs for the next five time steps and the type of the received cyberattack message. The label data $Y$ is defined as follows:
\begin{equation}
\begin{aligned}
Y=\{&Loc_{xi}, Loc_{yi},Atk_{i}\} \ i \in [1,5].
\end{aligned}
\end{equation}

Then, each onboard LSTM network accepts input data $X$ and generates the trajectory data for the next five steps and the received data's specific attack type. The objective function of the base LSTM model is to minimize the difference between the model output and the label data $Y$ of the future trajectory:
\begin{equation}
\begin{aligned}
\min O b j\left(\omega_{i}\right)=& \frac{1}{N}\sum_{n=1}^{5}\left|L o c_{p d t}^{n}-L o c_{lb}^{n}\right|^{2} \\
+&\frac{1}{N}\sum_{n=1}^{5}\left|A t k_{p d t}^{n}-A t k_{lb}^{n}\right|^{2},
\end{aligned}
\end{equation}
where $Loc_{pdt}$ and $Loc_{lb}$ refer to the trajectory prediction value and real trajectory data of the onboard LSTM network for the surrounding SVs, respectively. $Atk_{pdt}$ and $Atk_{lb}$ refer to the prediction of message attack type in the model and the real attack type in the ground truth.

After completing local training, each vehicle's model data can be transmitted to the Road Side Unit (RSU) cloud node through V2I for further FL. The further details are shown in Figure \ref{FDL}, which includes four steps: cyberattack detection and trajectory prediction, weights calculation, federated model aggregation, and model update. The specific progress of model aggregation is shown in Algorithm \ref{alg:FL}. The objective optimization function of the whole algorithm is to minimize the sum of each onboard model's objective optimization function:
\begin{equation}
\mathcal{P}: \min _{\{\omega_i\}} \ \sum_{i=1}^{N} Obj({\omega_i}),
\end{equation}
where $\omega_i$ refers to the onboard model of the $i^{th}$ vehicle, and $N$ refers to the total number of vehicles in the interest region. The detailed Model-robust Estimation (MrE) algorithm process will be further described in the next section.

\begin{algorithm}[!h]
    \caption{Federated Deep Learning Process of FL-TP}
    \label{alg:FL}
    \renewcommand{\algorithmicrequire}{\textbf{Input:}}
    \renewcommand{\algorithmicensure}{\textbf{Output:}}
    \begin{algorithmic}[1]
        \REQUIRE $\omega^{t}_n$, $dataBatch$
        \ENSURE $\omega$    
        
        \STATE  All vehicles complete model training and generate $\omega^{t}_n$
        \STATE  Generate random number $\gamma$

        \FOR{each round $t$ = 1, 2, \dots}
            \IF{$\gamma<0.2$}
                \FOR{each vehicle k}
                \STATE $\lambda_{n}^{t+1} \leftarrow {1/N}$
                \ENDFOR
            \ELSE
                \FOR{each vehicle k}
                \STATE $\lambda_{n}^{t+1} \leftarrow MrE(\omega^{t}_n, dataBatch)$\;
                
                \ENDFOR
            \ENDIF
            \STATE$\omega \leftarrow \sum_{n=1}^{N} \frac{1}{N} \lambda_{t+1}^{k} \omega_{t+1}^{k}$\;
            
        \ENDFOR
        \RETURN $Global Model \ \omega$
    \end{algorithmic}
\end{algorithm}

\section{Model-robust Estimation Algorithm}
Generally speaking, the onboard model that is trained with more good data in the training set can better predict trajectory and cyberattack type. To generate a greater weight $\lambda$ for the model trained with more good data, the MrE algorithm is proposed. Compared to the model trained with numerous cyberattacks, the model with a better effect is given a higher weight in the global model aggregation, which improves the convergence speed and effectiveness of the global model generated by the central server.
However, it is difficult to generate proper weights for the MrE algorithm because of the poor effectiveness of the onboard model in the first several episodes. To address this problem, an effectiveness judgment factor $\gamma$ is added to the FL-TP algorithm. The RSU will implement the federal average learning if the cyberattack detection accuracy of the global model is less than $\gamma$, otherwise, the MrE algorithm will be used. This paper sets the effectiveness judgment factor $\gamma$ as 0.2.
Moreover, different cyberattacks may cause different interference levels to the onboard model. For example, although the constant attack is considered to generate completely incorrect data, a tiny value offset attack can help the model avoid overfitting. Therefore, an influencing factor $\xi_{i}$ is also added to the MrE algorithm, which refers to the impact of the $i^{th}$ attack type. The MrE process is defined as follows:
\begin{equation}
\lambda_{i}=\frac{E_{i}}{\sum_{j=1}^{N} E_{j}} \quad E_{i}=\left(1-\sum_{i=1}^{K} \frac{U^{i}_{A t k} * \xi_{i}}{U_{ {total }}}\right),
\end{equation}
where $K$ represents the number of all types of attacks encountered in the LSTM model training, $U^{i}{Atk}$ and $U{total}$ represent the number of attack types $i$ in the training dataset, and the total number of training sets, $\xi_{i}$ is calculated based on the threats to the ego vehicle posed by different attack types by using the Beta function proposed in \cite{wang2021credibility}. 
To the best of our knowledge, there is no other research on the calculation of weights among all the different cyberattacks in the internet of vehicles.

        


        
            
        

\begin{table}[th]
  \centering
  \caption{Experimental Parameters}
  \label{tab:Experiment_Parameter}
  \begin{tabular}[l]{@{}lc}
  \toprule
      \textbf{Parameter} & \textbf{Value} \\
      \midrule
      \footnotesize Optimizer & \footnotesize SGD (Momentum:0.5) \\ 
      \footnotesize Time Length / Prediction Time Step & \footnotesize $10$ / $5$ \\ 
      \footnotesize Activation Function / Learning Rate  & \footnotesize $tanh$ / $10^{-5}$ \\ 
      
      \footnotesize Interest Region & \footnotesize $10km * 10km$ \\ 
      \footnotesize Vehicle Number/ Batch Size & \footnotesize $[4,20]$ / $128$ \\    
      \footnotesize Local Episode  / Global Episode & \footnotesize $10$  / $300$ \\ 
      \footnotesize Cyberattack Types / Penetration& \footnotesize $5$ / $[0.25,0.75]$ \\ 
    \bottomrule
  \end{tabular}
\end{table}

\begin{figure*}[htbp]
\centering
\subfigure[Prediction accuracy (cyberattack 25\%).]{
\begin{minipage}[t]{0.33\textwidth}
\centering
\includegraphics[width=\textwidth]{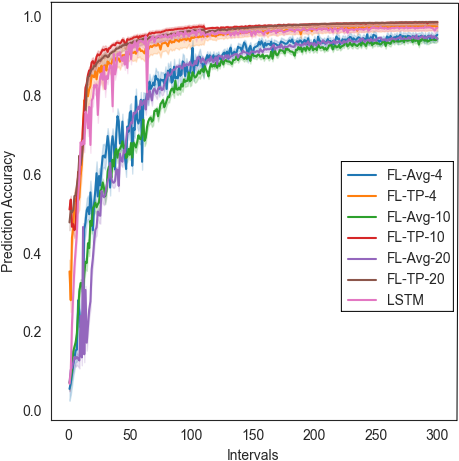}
\end{minipage}%
}%
\subfigure[Prediction accuracy (cyberattack 50\%).]{
\begin{minipage}[t]{0.336\textwidth}
\centering
\includegraphics[width=\textwidth]{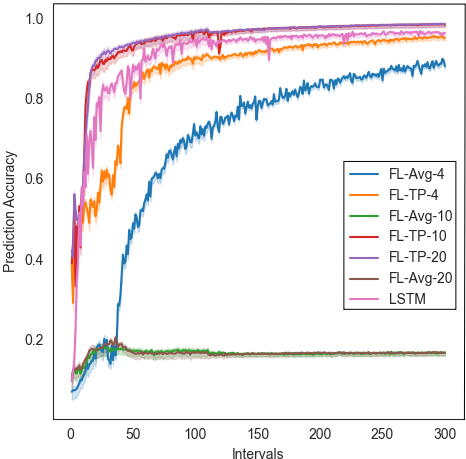}
\end{minipage}%
}%
\subfigure[Prediction accuracy (cyberattack 75\%).]{
\begin{minipage}[t]{0.332\textwidth}
\centering
\includegraphics[width=\textwidth]{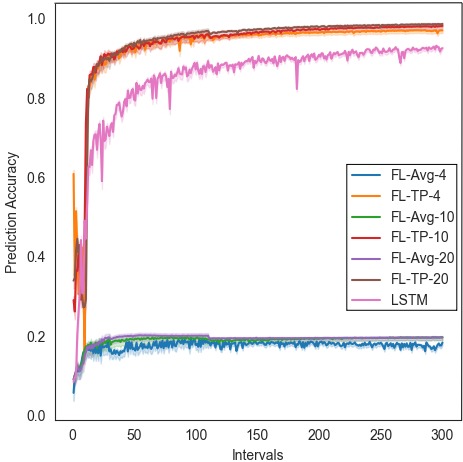}
\end{minipage}
}%

\subfigure[Prediction error (cyberattack 25\%).]{
\begin{minipage}[t]{0.33\textwidth}
\centering
\includegraphics[width=\textwidth]{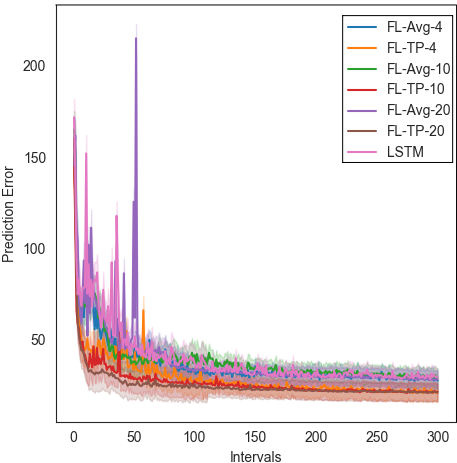}
\end{minipage}%
}%
\subfigure[Prediction error (cyberattack 50\%).]{
\begin{minipage}[t]{0.33\textwidth}
\centering
\includegraphics[width=\textwidth]{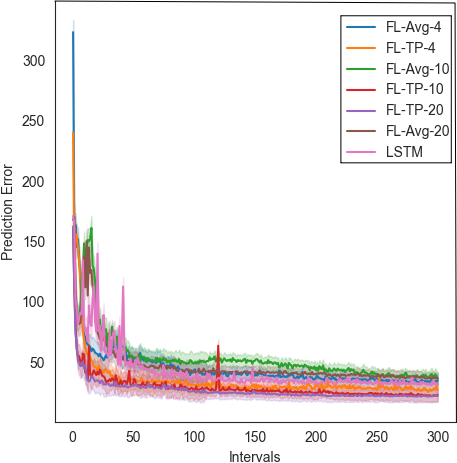}
\end{minipage}%
}%
\subfigure[Prediction accuracy (cyberattack 75\%).]{
\begin{minipage}[t]{0.33\textwidth}
\centering
\includegraphics[width=\textwidth]{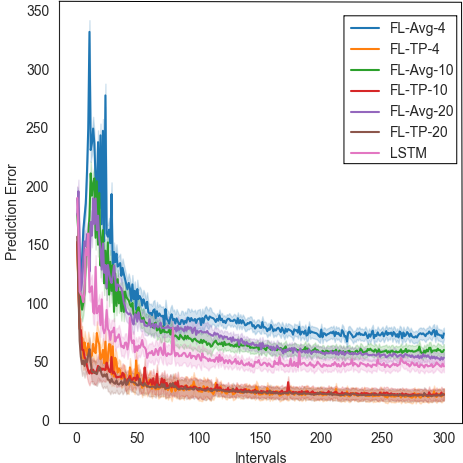}
\end{minipage}
}%
\centering
\caption{Experimental performance of proposed FL-TP algorithm}
\end{figure*}




\begin{table*}[t]
\setlength{\belowcaptionskip}{-0.2cm}   
\renewcommand\arraystretch{1.3}
\centering
\caption{Experimental Performance}
\label{tab:Experiment_Performance}
{
\begin{tabular}{c|ccc|ccc|ccc}
\hline
{\color[HTML]{333333} Attack level}                                                           & \multicolumn{3}{c|}{{\color[HTML]{333333} Cyberattack 25\%}}                                                                                                                                                                                        & \multicolumn{3}{c|}{{\color[HTML]{333333} Cyberattack 50\%}}                                                                                                                                                                                        & \multicolumn{3}{c}{{\color[HTML]{333333} Cyberattack75\%}}                                                                                                                                                                                          \\ \hline
{\color[HTML]{333333} Method}                                                                 & {\color[HTML]{333333} FL-TP-4}                                                  & {\color[HTML]{333333} FL-TP-10}                                                 & {\color[HTML]{333333} FL-TP-20}                                                 & {\color[HTML]{333333} FL-TP-4}                                                  & {\color[HTML]{333333} FL-TP-10}                                                 & {\color[HTML]{333333} FL-TP-20}                                                 & {\color[HTML]{333333} FL-TP-4}                                                  & {\color[HTML]{333333} FL-TP-10}                                                 & {\color[HTML]{333333} FL-TP-20}                                                 \\ \hline
{\color[HTML]{333333} \begin{tabular}[c]{@{}c@{}}Prediction  accuracy\end{tabular}}          & {\color[HTML]{333333} \begin{tabular}[c]{@{}c@{}}0.974\\  $\pm$ 0.0008\end{tabular}} & {\color[HTML]{333333} \begin{tabular}[c]{@{}c@{}}0.978\\  $\pm$ 0.0007\end{tabular}} & {\color[HTML]{333333} \begin{tabular}[c]{@{}c@{}}0.978\\  $\pm$ 0.0008\end{tabular}} & {\color[HTML]{333333} \begin{tabular}[c]{@{}c@{}}0.959\\  $\pm$ 0.003\end{tabular}}  & {\color[HTML]{333333} \begin{tabular}[c]{@{}c@{}}0.977\\  $\pm$ 0.0008\end{tabular}} & {\color[HTML]{333333} \begin{tabular}[c]{@{}c@{}}0.977\\  $\pm$ 0.001\end{tabular}}  & {\color[HTML]{333333} \begin{tabular}[c]{@{}c@{}}0.963\\  $\pm$ 0.001\end{tabular}}  & {\color[HTML]{333333} \begin{tabular}[c]{@{}c@{}}0.973\\  $\pm$ 0.001\end{tabular}}  & {\color[HTML]{333333} \begin{tabular}[c]{@{}c@{}}0.979\\  $\pm$ 0.001\end{tabular}}  \\ \hline
{\color[HTML]{333333} \begin{tabular}[c]{@{}c@{}}Prediction  error\end{tabular}}             & {\color[HTML]{333333} \begin{tabular}[c]{@{}c@{}}21.888\\  $\pm$ 0.736\end{tabular}} & {\color[HTML]{333333} \begin{tabular}[c]{@{}c@{}}21.607\\  $\pm$ 0.429\end{tabular}} & {\color[HTML]{333333} \begin{tabular}[c]{@{}c@{}}21.476\\  $\pm$ 0.297\end{tabular}} & {\color[HTML]{333333} \begin{tabular}[c]{@{}c@{}}28.386\\  $\pm$ 1.791\end{tabular}} & {\color[HTML]{333333} \begin{tabular}[c]{@{}c@{}}23.149\\  $\pm$ 0.592\end{tabular}} & {\color[HTML]{333333} \begin{tabular}[c]{@{}c@{}}22.083\\  $\pm$ 0.312\end{tabular}} & {\color[HTML]{333333} \begin{tabular}[c]{@{}c@{}}20.874\\  $\pm$ 1.060\end{tabular}} & {\color[HTML]{333333} \begin{tabular}[c]{@{}c@{}}22.254\\  $\pm$ 0.678\end{tabular}} & {\color[HTML]{333333} \begin{tabular}[c]{@{}c@{}}21.447\\  $\pm$ 0.480\end{tabular}} \\ \hline
{\color[HTML]{333333} Method}                                                                 & {\color[HTML]{333333} FL-Avg-4}                                                 & {\color[HTML]{333333} FL-Avg-10}                                                & {\color[HTML]{333333} FL-Avg-20}                                                & {\color[HTML]{333333} FL-Avg-4}                                                 & {\color[HTML]{333333} FL-Avg-10}                                                & {\color[HTML]{333333} FL-Avg-20}                                                & {\color[HTML]{333333} FL-Avg-4}                                                 & {\color[HTML]{333333} FL-Avg-10}                                                & {\color[HTML]{333333} FL-Avg-20}                                                \\ \hline
{\color[HTML]{333333} \begin{tabular}[c]{@{}c@{}}Prediction accuracy\end{tabular}}          & {\color[HTML]{333333} \begin{tabular}[c]{@{}c@{}}0.946\\  $\pm$ 0.005\end{tabular}}  & {\color[HTML]{333333} \begin{tabular}[c]{@{}c@{}}0.931\\  $\pm$ 0.0047\end{tabular}} & {\color[HTML]{333333} \begin{tabular}[c]{@{}c@{}}0.938\\  $\pm$ 0.003\end{tabular}}  & {\color[HTML]{333333} \begin{tabular}[c]{@{}c@{}}0.871\\  $\pm$ 0.009\end{tabular}}  & {\color[HTML]{333333} \begin{tabular}[c]{@{}c@{}}0.160\\  $\pm$ 0.001\end{tabular}}  & {\color[HTML]{333333} \begin{tabular}[c]{@{}c@{}}0.160\\  $\pm$ 0.0009\end{tabular}} & {\color[HTML]{333333} \begin{tabular}[c]{@{}c@{}}0.170\\  $\pm$ 0.006\end{tabular}}  & {\color[HTML]{333333} \begin{tabular}[c]{@{}c@{}}0.190\\  $\pm$ 0.0006\end{tabular}} & {\color[HTML]{333333} \begin{tabular}[c]{@{}c@{}}0.190\\  $\pm$ 0.0003\end{tabular}} \\ \hline
{\color[HTML]{333333} \begin{tabular}[c]{@{}c@{}}Prediction  error\end{tabular}}             & {\color[HTML]{333333} \begin{tabular}[c]{@{}c@{}}29.211\\  $\pm$ 1.059\end{tabular}} & {\color[HTML]{333333} \begin{tabular}[c]{@{}c@{}}30.219\\  $\pm$ 0.803\end{tabular}} & {\color[HTML]{333333} \begin{tabular}[c]{@{}c@{}}28.766\\  $\pm$ 0.463\end{tabular}} & {\color[HTML]{333333} \begin{tabular}[c]{@{}c@{}}34.403\\  $\pm$ 0.722\end{tabular}} & {\color[HTML]{333333} \begin{tabular}[c]{@{}c@{}}38.722\\  $\pm$ 1.138\end{tabular}} & {\color[HTML]{333333} \begin{tabular}[c]{@{}c@{}}37.867\\  $\pm$ 0.821\end{tabular}} & {\color[HTML]{333333} \begin{tabular}[c]{@{}c@{}}73.450\\  $\pm$ 1.729\end{tabular}} & {\color[HTML]{333333} \begin{tabular}[c]{@{}c@{}}59.235\\  $\pm$ 1.222\end{tabular}} & {\color[HTML]{333333} \begin{tabular}[c]{@{}c@{}}54.682\\  $\pm$ 1.289\end{tabular}} \\ \hline
{\color[HTML]{000000} Method}                                                                 & \multicolumn{3}{c|}{{\color[HTML]{000000} LSTM}}                                                                                                                                                                                                    & \multicolumn{3}{c|}{{\color[HTML]{000000} LSTM}}                                                                                                                                                                                                    & \multicolumn{3}{c}{{\color[HTML]{000000} LSTM}}                                                                                                                                                                                                     \\ \hline
{\color[HTML]{000000} \begin{tabular}[c]{@{}c@{}}Prediction  accuracy\end{tabular}}          & \multicolumn{3}{c|}{{\color[HTML]{000000} 0.964 $\pm$ 0.002}}                                                                                                                                                                                            & \multicolumn{3}{c|}{{\color[HTML]{000000} 0.956 $\pm$ 0.003}}                                                                                                                                                                                            & \multicolumn{3}{c}{{\color[HTML]{000000} 0.915 $\pm$ 0.007}}                                                                                                                                                                                             \\ \hline
{\color[HTML]{000000} \begin{tabular}[c]{@{}c@{}}Prediction error\end{tabular}}             & \multicolumn{3}{c|}{{\color[HTML]{000000} 29.947 $\pm$ 0.989}}                                                                                                                                                                                           & \multicolumn{3}{c|}{{\color[HTML]{000000} 32.650 $\pm$ 1.391}}                                                                                                                                                                                           & \multicolumn{3}{c}{{\color[HTML]{000000} 47.510 $\pm$ 1.541}}                                                                                                                                                                                            \\ \hline
{\color[HTML]{000000} \begin{tabular}[c]{@{}c@{}}Improvement\\ (Accuracy min)\end{tabular}} & {\color[HTML]{000000} 1.03\%}                                                   & {\color[HTML]{000000} 1.40\%}                                                   & {\color[HTML]{000000} 1.40\%}                                                   & {\color[HTML]{000000} 0.31\%}                                                   & {\color[HTML]{000000} 2.19\%}                                                   & {\color[HTML]{000000} 2.19\%}                                                   & {\color[HTML]{000000} 5.21\%}                                                   & {\color[HTML]{000000} 6.31\%}                                                   & {\color[HTML]{000000} 6.99\%}                                                   \\ \hline
\begin{tabular}[c]{@{}c@{}}Improvement\\ (Error min)\end{tabular}                           & 25.06\%                                                                         & 27.84\%                                                                         & 25.34\%                                                                         & 13.05\%                                                                         & 29.09\%                                                                         & 32.36\%                                                                         & 56.06\%                                                                         & 53.16\%                                                                         & 54.86\%                                                                         \\ \hline

\end{tabular}}
\end{table*}

\section{Performance Evaluation}
This section introduces the dataset used, experiment setups, and the performance of the proposed FL-TP algorithm. The experiments are implemented under three different traffic scenarios to prove the reliability and robustness of the proposed FL-TP algorithm. All source code about the algorithm, data generation, and performance evaluation in this paper are provided on GitHub (https://github.com/CoderTylor/FL-TP).

\subsection{VeReMi trajectory dataset}\label{AA}
The VeReMi dataset is generated from a simulation environment in LuST scenarios with the SUMO and VEINS simulators. It is being used in experiments to verify the effectiveness of the FL-TP algorithm under cyberattacks. The VeReMi contains five attack types and one real vehicle data type, including GPS data about the local vehicle and Basic Safety Message (BSM) messages from surrounding vehicles. The bogus message types contain constant, constant offset, random, random offset, and eventual stop.

\subsection{Experiment setup}\label{BB}

Inspired by the memory-keeping ability of the LSTM model, an LSTM-centered model is trained as an onboard vehicle training task to predict the vehicle's trajectory and attack-type of received messages in the next five time steps. SGD optimizer with a momentum value of  0.5 is used, and the learning rate is set as $10^{-5}$. The number of total SVs ranges from 4 to 20, while the attack penetration is ranged from 0.25 to 0.75. The complete onboard model and experiment parameters are listed in Table \ref{tab:Experiment_Parameter}. To the best of our knowledge, there is no research investigating the trajectory prediction algorithm under cyberattack scenarios. Two other baseline algorithms are added as comparison schemes to clarify the performance of FL-TP better, which are:

(1) Centralized LSTM-based trajectory prediction algorithm (LSTM): Based on the centralized LSTM model to predict future highway vehicle trajectories proposed in \cite{8317913}. The cyberattack type is added as output, and all the given parameters of the VeRemi dataset are defined as input in the code. 

(2) Federated-Average learning (Fed-Avg): The centralized LSTM model is modified as a distributed algorithm based on the open-source code of federated average learning framework proposed in \cite{mcmahan2017communication}. 

The FL-TP algorithm is trained under three different traffic scenarios and reported the average performance statistics (mean and standard deviation) across fifty test episodes in the subsequent discussion. Accordingly, the performance of each baseline method is also averaged across fifty test episodes on the same traffic scenarios.

Two indicators are chosen to comprehensively evaluate the performance of the proposed FL-TP algorithm, which are trajectory prediction error and cyberattacks prediction accuracy. 
The prediction error is used to evaluate the trajectory predictability of the onboard model. It is defined as the average Euclidean distance between the predicted trajectory and actual trajectory in the VeRemi dataset, which is:
\begin{equation}
\begin{aligned}
Pr&ediction \\
&error
\end{aligned}=\frac{1}{N} \sum_{i=1}^{N} \sqrt{\left(x_{p d t}^{i}-x_{l b}^{i}\right)^{2}+\left(y_{p d t}^{i}-y_{l b}^{i}\right)^{2}}.
\end{equation}

Prediction accuracy is used to evaluate the accuracy of cyberattack judgment. It is calculated as the ratio of the true attack judgment number to the total number of samples in the test dataset, which is defined as:
\begin{equation}
 Prediction \ accuracy =\frac{T J}{T J+F J},
\end{equation}
where $TJ$ and $FJ$ refer to the number of true judgment and false judgment generated by the onboard model. True judgment means that the error between the predicted attack ID and the actual attack ID is less than the decision threshold $\eta$ and vice versa. The judgment of model output is defined as follows:
\begin{equation}
Attack \ judgment =\left\{\begin{array}{ll}
|ATK_{pdt}-ATK_{lb}|<\eta \ \text { True }\\
|ATK_{pdt}-ATK_{lb}|>\eta \ \text { False },
\end{array} \quad\right. 
\end{equation}
where $ATK_ {pdt} $ and $ATK_ {lb} $ refer to the attack type judgment of the onboard model and attack type label. The parameter $\eta$ refers to the decision threshold, and it is set to $0.5$. 

\subsection{Performance of the FL-TP algorithm}\label{CC}

Figures 3(a), 3(b), and 3(c) successively illustrate the cyberattack prediction accuracy of the FL-TP algorithm, FL-Avg algorithm, and the centralized LSTM algorithm under three different attack penetration rates and different traffic scenarios.
It can be seen that FL-TP has lower trajectory prediction error and higher cyberattack prediction accuracy than other baseline algorithms in all scenarios in the 200-300 episodes. In addition, compared with the other two baseline algorithms, FL-TP can achieve a rapid convergence of the model in the cyberattack attack penetration scenario in the first 100 episodes.
The FL-Avg algorithm can no longer achieve cyberattack detection when the attack penetration rate exceeds 50\%. 
In medium and high-density scenarios, Fed-Avg's prediction accuracy of cyberattacks reaches 0.17\% on average. Although the centralized LSTM algorithm can distinguish the cyberattack type well under all penetration of attacks, it still has the problem of slow model convergence because of numerous training data in merely one agent. The average cyberattack prediction accuracy of FL-TP is 3\% higher than that of the centralized LSTM algorithm.

Figures 3(d), 3(e), and 3(f) successively illustrate the trajectory prediction error of the FL-TP algorithm and the other baseline algorithms under three different attack penetration rates with varying scenarios of traffic.
Although FL-TP and the other two baseline algorithms can achieve model convergence under the three attack levels, FL-TP can achieve a lower trajectory prediction error among all algorithms in the 200-300 episodes. Moreover, trajectory prediction error will have a higher fluctuation value in the model training process with the increase of attack permeability. 
The specific experiment results are shown in the Table \ref{tab:Experiment_Performance}.

\section{Conclusion}
This paper proposed a FL-TP algorithm. 
to provide accurate and real-time prediction in trajectory and cyberattack with five kinds of cyberattack penetration. The algorithm is evaluated with a publicly available dataset VeRemi and demonstrated its outstanding performance. 
Our future work will concentrate on carrying out extensive simulations using a large-scale realistic scenario to further assess the proposed algorithm's performance.

\bibliographystyle{elsarticle-num}
\small
\bibliography{egbib}

\begin{thebibliography}{10}
\expandafter\ifx\csname url\endcsname\relax
  \def\url#1{\texttt{#1}}\fi
\expandafter\ifx\csname urlprefix\endcsname\relax\def\urlprefix{URL }\fi
\expandafter\ifx\csname href\endcsname\relax
  \def\href#1#2{#2} \def\path#1{#1}\fi

\bibitem{naseer2019swarm}
A.~Naseer, M.~Jaber, Swarm wisdom for smart mobility: The next generation of
  autonomous vehicles, in: Proc. 2019 IEEE SmartWorld, Ubiquitous Intelligence
  \& Computing, Advanced \& Trusted Computing, Scalable Computing \&
  Communications, Cloud \& Big Data Computing, Internet of People and Smart
  City Innovation, IEEE, 2019, pp. 1943--1949.

\bibitem{9650522}
M.~H. Bahonar, M.~J. Omidi, H.~Yanikomeroglu, Low-complexity resource
  allocation for dense cellular v2x comms., IEEE Open J. Commun. Soc. 2 (2021)
  2695--2713.

\bibitem{9839627}
Y.~Wang, J.~Zhou, G.~Feng, X.~Niu, S.~Qin, Blockchain assisted fed. learning
  for enabling net. edge intel., IEEE Net. (2022) 1--7.

\bibitem{guler2021framework}
B.~Güler, A.~Yener, A framework for sustainable fed. learning, in: 2021 19th
  Int. Symp. on Modeling and Opt. in Mobile, Ad hoc, and Wireless Net. (WiOpt),
  IEEE, 2021, pp. 1--8.

\bibitem{alvi2022utility}
S.~A. Alvi, Y.~Hong, S.~Durrani, Utility fairness for diff. private fed.
  learning-based wireless iot net., IEEE Internet of Things J. (2022).

\bibitem{jiang2022multi}
M.~Jiang, T.~Wu, Z.~Wang, Y.~Gong, L.~Zhang, R.~P. Liu, A multi-intersection
  vehicular cooperative control based on end-edge-cloud computing, IEEE Trans.
  Veh. Technol. 71~(3) (2022) 2459--2471.

\bibitem{yuan2023hierarchical}
Z.~Yuan, Z.~Wang, X.~Li, L.~Li, L.~Zhang, Hierarchical trajectory planning for
  narrow-space automated parking with deep reinforcement learning: A federated
  learning scheme, Sensors 23~(8) (2023) 4087.

\bibitem{9417511}
M.~Karimzadeh, R.~Aebi, A.~M.~d. Souza, Z.~Zhao, T.~Braun, S.~Sargento,
  L.~Villas, Reinforcement learning-designed lstm for trajectory and traffic
  flow prediction, in: 2021 IEEE Wireless Commun. and Netw. Conf. (WCNC), 2021,
  pp. 1--6.

\bibitem{8933492}
Y.~Xing, C.~Lv, D.~Cao, Personalized vehicle trajectory prediction based on
  joint time-series modeling for connected vehicles, IEEE Trans. Veh. Technol.
  69~(2) (2020) 1341--1352.

\bibitem{belenko2018synthetic}
V.~Belenko, V.~Krundyshev, M.~Kalinin, Synthetic datasets generation for
  intrusion detection in vanet, in: Proc. 11th Int. Conf. Secur. Inf. Netw.,
  2018, pp. 1--6.

\bibitem{9120761}
T.~V. Khoa, Y.~M. Saputra, D.~T. Hoang, N.~L. Trung, D.~Nguyen, N.~V. Ha,
  E.~Dutkiewicz, Collaborative learning model for cyberattack detection systems
  in iot industry 4.0, in: Proc. 2020 IEEE Wirel. Commun. Netw. Conf. (WCNC),
  2020, pp. 1--6.

\bibitem{wang2021credibility}
Z.~Wang, X.~Li, T.~Wu, C.~Xu, L.~Zhang, A credibility-aware swarm-federated
  deep learning framework in internet of vehicles, Digit. Commun. Netw. (2023).

\bibitem{8317913}
F.~Altché, A.~de~La~Fortelle, An lstm network for highway trajectory
  prediction, in: 2017 IEEE 20th Int. Conf. on Intell. Transp. Syst. (ITSC),
  IEEE, 2017, pp. 353--359.

\bibitem{mcmahan2017communication}
B.~McMahan, E.~Moore, D.~Ramage, S.~Hampson, B.~y~Arcas,
  Communication-efficient learning of deep networks from decentralized data,
  in: Artif. Intell. and Stat., PMLR, 2017, pp. 1273--1282.

\end{thebibliography}
\end{document}